# Scenario and Sensitivity Analysis for Flooding Vulnerability using Genetic Algorithms

Vena Pearl Boñgolan[a]
vabongolan@dcs.upd.edu.ph

Oreste Terranova.[b]
terranova@irpi.cnr.it

Edward Nataniel Apostol[a]
edward.nataniel@yahoo.com

Joshua Kevin Cruz [a]
joshuakevina.cuz@yahoo.com

a) The Scientific Computing Laboratory

Department of Computer Science, College of Engineering,

University of the Philippines Diliman

b) CNR-IRPI Cosenza, Italy

## ABSTRACT

We try to answer the question: "can we 'modify' our neighborhoods to make them less vulnerable to flooding?" We minimize flooding vulnerability for a city in the central plain of Luzon, by modeling the city as a biological organism with 'traits', and try to 'breed' a 'champion' city (with a low flooding vulnerability) via a genetic algorithm. The result is a description of the traits the *barangays* (neighborhoods) should have (the 'design' of the city). As far as we can tell, this kind of modeling has not been attempted before. The different components of flooding vulnerability were investigated, and each was given a weight, which allows us to express vulnerability as a weighted sum; this serves as the fitness function for the genetic algorithm. We also allowed non-linear interactions among related but independent components, viz, poverty and mortality rate, and literacy and radio/TV penetration. The two-table system we used to prioritize the components of vulnerability is prone to subjectivity, a common problem in analyses of vulnerability. Thus, a sensitivity analysis was done, which gave a design with a 24% decrease in vulnerability alongside a 14% percent decrease in cost, a significant improvement over this initial scenario analysis, where the proposed design had a 12% decrease in vulnerability with a one percent increase in cost.

## Categories and Subject Descriptors

I.2.8 [**Problem Solving, Control Methods, and Search**] genetic algorithms, multi-objective optimization.

## General Terms

Algorithms

## Keywords

Risk Assessment, Vulnerability

## INTRODUCTION

Because of its geographic location in both the Southeast Asian monsoon and typhoon belts, the Philippines is prone to flooding in various areas. This disaster has affected the second highest number of people in the Philippines from 1900 to 2014, storm being the first. It has been taking lives and damaging properties annually.

One city that experiences frequent flooding is the city of Urdaneta in Pangasinan, Philippines (City of Urdaneta 2012). A second class city located in East-Central Pangasinan, situated on 15o 56' to 16o 0' latitude and 120o 29' to 120o 37' longitude, about 186 km north of Manila, with an area of 100.26 km$^2$ (38.71 sq mi, per the Department of Environment and Natural Resources, Philippines), and a population of 125,451 (2010 Census, National Statistics Office, Philippines). Figure 1 shows the satellite photo. It is a very prosperous agricultural town in the central plain of Luzon, an area called the 'rice granary' of the country, with an infant mortality of 1.2/ 1000 in 2012, when the national average was 30/1000 (World Bank, 2016). It is earmarked for development ( a 'growth pole' of the Province of Pangasinan), but still does not have the problems of a major city (68.61% urbanization level, population growth rate of 1.10%, population density of 10 persons/ has). It is fast becoming a business and trading center, home of the largest fruits and vegetables wholesale market in Northern Luson, and is one of the biggest livestock markets in the country. The city experiences seasonal floods especially *barangays* along the river north of the city (Musing/ Macalong River). City planners are challenged to come up with preparedness plans for flood and other disasters. In the older studies done by Bongolan et al. (2013, 2015), they modeled a city as an organism with chromosomes to define its traits, and used a genetic algorithm to come up with an optimal arrangement of *barangays* based on their traits, aiming to lessen flood risk. There, they introduced the chromosomes which were input into the genetic algorithm (area's urbanized area ratio, literacy rate, mortality rate, percent of population under poverty, radio/TV penetration and the state of structural and non-structural measures) (Bongolan et at. 2013); a weighted vulnerability function (Bongolan et a. 2015), wherein the weights were determined by a two-table analysis (State of Michigan, USA, Vulnerability Assessment Protocol ;see URL in reference section). They also introduced an assumed cost function, which might be taken as a penalty for moving out of vulnerability. Both works applied the genetic algorithm on a hypothetical city, which we now extend into a real city. Interactions between independent, but nevertheless related components like poverty and mortality were allowed (Hall et al. 2003; Huang et al. 2010; Scheuer et al. 2013). While one is usually expected to be high when the other is high, these issues are, however, addressed separately by different government agencies. We take the mortality chromosome to be an indicator of the overall health of our population, thus agencies like the Department of Health could address concerns here. On the other hand, poverty is more complex, and most likely to be addressed by several government agencies, including (in the Philippine setting): The National Economic and Development Authority (NEDA) and the National Anti-Poverty Commission. Similarly, literacy rate and radio/ TV penetration are related, because they both address disaster preparedness and information, but again, might be handled by different government and private agencies.

However, the previous works did not address the practical cost of rearranging a city, like the cost of relocation. This is addressed in the current study by adding chromosomes that address exposure.

An immediate criticism earlier works received was the subjectivity inherent in making judgement calls concerning an area, the population therein, and it possible response to disasters. Scenario and sensitivity analyses were added to identify the flood risk in *barangays* being studied using risk factors (the chromosomes) (Elmoustafa 2012), but its best use is to make the analysis 'objective', by coming up with a set of weights not of our choosing, but still has the effect of minimizing the over-all vulnerability and cost, thus improving the fitness function.

## Background on Genetic Algorithms

Following natural evolution, i.e., inheritance, mutation, selection and crossover, Genetic Algorithm (GA) has gained wide acceptance in solving real-world engineering optimization problems (Kingston 2011). Similar to dog or cattle breeding, the objective is to 'breed' something, e.g., a city, or a structure, with certain desirable traits. In the field of flood management, GAs (genetic algorithm) have been used in the design of flood control structures (Wallace & Louis 2003), model calibration (Lan 2001), flood plain management (Karamouz et al. 2009), management of precipitation uncertainty in models (Maskey et al. 2004), flood forecasting (Mukerji et al. 2009), and dynamic control of reservoir operations (Li et al. 2010).

GA was used because of this case's high dimensionality similar to a multi-objective urban planning problem tackled by Balling et al. (1999) where direct methods are intractable, if not impossible to use. We settled on GA inspite of the well-known drawbacks of heuristic methods, which is: they are good in finding a solution, but most likely not the best solution or the global minumum or maximum.

## METHODOLOGY

We model the city of Urdaneta as a collection of organisms called *barangays,* each has a set of chromosomes describing its 'traits', and we will 'breed' a city with the desirable trait of having a low flooding vulnerability and cost. This will be achieved by applying a genetic algorithm to define the traits each *barangay* should have, which would result in the lowest over-all vulnerability for the city. Sixteen *barangays* of Urdaneta were selected for this study, as these have been identified as the most flood prone, per the city profile (City of Urdaneta, 2012).

First, we define the physical area, Figure 2, which was coloured from physical properties of the area which make it more or less vulnerable to flooding, and will be used to multiply the area's vulnerability. A river runs through the north of the city, and following the city profile (Urdaneta, 2012), risk factors were assigned per Figure 2. Most at risk have a factor of two, (in red); less prone to flooding have a factor of one (pink), and even less prone have a factor of ½, (light pink).

We define the chromosomes for the organisms/ *barangays* as binary numbers ("11" means 3 decimal, "00" means zero, so they range from zero to three). This will later feature in the calculation of cost below.

1) <u>Urbanized area ratio</u> - The less people and infrastructure that can be affected, the lower the vulnerability.

  11 - Highly urbanized

  10 - Moderately urbanized

  01 - A little urbanized

  00 - Not urbanized

2) <u>Literacy Rate</u> - Requires that communities understand warning signs from the government and follow instructions from emergency responders.

  11 - More than 75% are illiterate

  10 - 50 - 75% are illiterate

  01 - 25 - 50% are illiterate

  00 - 0 - 25% are illiterate

We similarly define:

3) Mortality Rate - An indicator of the general health of the population.

4) Population under poverty

5) TV / radio penetration rate - Aside from informing the public of danger, may also be used to train and prepare for disaster.

6) State of non-structural measures - Laws legislated, and estimated compliance with those laws, e.g., respecting easements on waterways, cleaning drainages, litter prevention, etc.

7) State of structural measures - Infrastructure built to prevent and/or control floods, i.e. drainage, flood gates, pumping stations.

The following static chromosomes concern exposure rather than vulnerability; added to determine the cost of implementing a design, and are not allowed to change over the generations.

8) Percentage of population - The higher the percentage of population in a specific *barangay*, the greater the exposure and risk.

   11 - Percentage is very high
   10 - Percentage is high
   01 - Percentage is average
   00 - Percentage is low

9) Percentage of area / extent - The larger the extent, more infrastructure and residents are more likely to be affected.

   11 - Percentage of area is very large
   10 - Percentage of area is large
   01 - Percentage of area is average
   00 - Percentage of area is small

We similarly define:

10) Economic Value - The higher the economic value, the greater the exposure.

11) Cost of Relocation - The higher the cost of relocation, the higher the exposure.

Vulnerability is assumed to be a weighted sum of its components, and we also allow for nonlinear interactions between mortality rate and poverty; and literacy rate and TV/radio penetration, to put some synchronicity in the design.

$$V_i = S_i * (W_{Urbanized}*X_{Urbanized} +$$
$$W_{Literacy}*X_{Literacy}*W_{TvRadio}*X_{TvRadio} + W_{Mortality}*X_{Mortality}*$$
$$W_{Poverty}*X_{Poverty} + W_{Nonstructural}*X_{Nonstructural} +$$
$$W_{Structural}*X_{Structural} + W_{Population}*X_{Population} +$$
$$W_{Extent}*X_{Extent} + W_{EconomicValue}*X_{EconomicValue} +$$
$$W_{CostOfRelocation} * X_{CostOfRelocation})$$

Where:

$V_i$ = vulnerability of the ith cell/*barangay*

$S_i$ = vulnerability factor of the ith cell/barangay (Figure 2)

$W_c$ = weight of chromosome c

$X_c$ = value of chromosome c (00 to 11 in binary)

The genetic algortihm searches the space of the X variables above, which define the traits of each *barangay*. The weights W above will be calculated and refined via a sensitivity analysis, with experiments which take maximum, minimum and median values from the initial set of weights. We now define the cost function or penalty function in this multi-objective optimization, to counter-act vulnerability: a small vulnerability comes with a high cost. We first take the three's complements of the variables, e.g., three minus mortality rate. The complement was then used to compute for the cost of improving the *barangay*'s mortality rate, because the lower the value of mortality rate, the bigger the expenditure on health. We enter the complements into an exponential, quadratic or linear function, depending on whether or not a solution is expensive (like poverty alleviation), relatively expensive (like improving health) or inexpensive, like improving literacy.

We assume exponential growth for 'expensive' activities like urbanization, poverty alleviation and building structural measures. All other chromosomes are assumed to have linear penalties (inexpensive), except the mortality variable, which is assumed to be quadratic (doctors affordable, medication costly, at least in the Philippines). These cost functions are currently hypotheses, and could be interpolated/ inferred from data, when and where available. This could be the object of future research.

$$C_i = \exp(3 - X_{Urbanized}) + (3 - X_{Literacy}) + (3 - X_{Mortality})^2 + \exp(3 - X_{Poverty}) + (3 - X_{TVRadio}) + (3 - X_{Nonstructural}) + \exp(3 - X_{Structural}) + (3 - X_{Population}) + \exp(3 - X_{EconomicValue}) + \exp(3 - X_{CostOfRelocation}))/S_i$$

Where:

$C_i$ = cost of the ith cell/barangay

$S_i$ = vulnerability factor of the ith cell/barangay (Figure 2)

$X_c$ = value of chromosome c ($00_2$ to $11_2$)

The initial values of the chromosomes are assigned based on Urdaneta's city profile, and input into MatLab's genetic algorithms toolbox, together with the vulnerability and cost functions.

Finally, we determine the weights to be used in the vulnerability function. Tables 1 and 2 come from the State of Michigan, USA, Vulnerability Assessment Protocol (see URL in reference section), originally used for prioritizing hazards, which we now use to prioritize the components of vulnerability. Instead of 'Hazard Aspect', we say 'Flooding', as we answer the questions, to come up with the set of interrogators. We took only the first two columns, those under "always very important" and "usually important" in coming up with Table 2 (only a portion is shown). This, and the succeeding step, is subjective, and will be improved with a sensitivity analysis, to be described later in the section.

We now try to distribute 100 points among the four marked "always very important" and the three marked "usually important", in such a way that each aspect in the first column weighs at least twice as the aspects in the second column, while keeping them whole numbers, for ease of later calculations (sensitivity analysis). We started with the first four aspects (always very important) with percentages of 18 or 19 points each while the last three aspects have eight or nine points each. An alternative method might be the analytic hierarchy process done by Roy and Blaschke (2013). We find the two-table process used by the State of Michigan (2011) and Einarsson and Rausand (1998) more suitable for this problem.

We now construct Table 2 by interrogating each of the eleven components of vulnerability (listed vertically) with the seven aspects/ interrogators we chose (listed horizontally). We gave ratings to each component based on how it affects the specific aspect, with 1 being the lowest and 10 being the highest. A sample calculation for Urbanization would be

9*.19+ 7*.19 + 6*.19 + 4*.09+5*.18+6*.08+2*.08 = 6.08.
.

Urbanization a score of "9" for the aspect "Capacity to cause physical damage", which has 19%, hence the calculation "9*.19". The resulting sum of 6.08 is the "weight" of Urbanization.

Table 2 shows our table analysis, and the resulting weights:

Urbanization: 6.08

Literacy: 3.88

Mortality: 4.72

Poverty: 4.02

TV/radio Penetration: 5.49

Non- Structural Measures: 3.78

Structural Measures: 6.76

Percentage of population: 6.09

Percentage of area/extent: 3.29

Economic value: 5.43

Cost of relocation: 4.86

Our previous work (Bongolan et al. 2013 and 2015) did not specify a study area, but an urban area like Metro Manila was in fact the model. There, "urbanization" came up as the most important component of vulnerability, but in Urdaneta, which is still largely agricultural, "structural measures" came up as the most important component.

We entered an initial set of chromosomes for each barangay, based on the city profile. From the formulae given above, initial vulnerability was 18.0137, and initial cost was 872.006. The genetic algorithm gave a design whose vulnerability decreased to 15.7864 (12% decrease) and cost increased to 879.1572 (1% increase).

Finally, a sensitivity analysis was done, comparing the results of running the GA using different sets of weights. Here, we probe the weights being used (not the chromosomes or traits of the city), trying to come up with a non-subjective set of weights that will still achieve our goals of minimizing vulnerability and cost. The sets of weights were produced by minimizing, maximizing, or setting to a median the percentage of a hazard aspect. From Table 2, the original percentages of the aspects are {0.19, 0.19, 0.19, 0.09, 0.18, 0.08, 0.08} respectively.

As an example, take the "Size of Affected Area", currently eight percent: minimizing "0.08" to "0.01" and redistributing the 0.07 percent over the six other aspects (0.01167) gives us new percentages {0.20167, 0.20167, 0.20167, 0.10167, 0.19167, 0.09167, 0.01}. Using these new percentages for Urbanization, we get:

9*0.20167 + 7*0.20167 + 6*0.20167 + 4*0.10167 + 5*0.19167 + 6*0.09167 + 2*0.01 = 6.37.

This weight shows in Table 3, Size of Affected Area (last row), minimize > $M_1$ (third column). The median and maximum percentages for this aspect are 0.28 and 0.55 respectively. The minimum, median, and maximum percentage for any aspect is calculated by adding (-)0.07, 0.2, and 0.47 to its original percentage. The value (-0.07) was chosen so that the smallest aspect percentage (Size of affected area, originally 0.08), will still be positive, i.e., we chose the maximum number we can subtract from all the aspects so that all percentages will still be positive. Similarly, 0.47 is the maximum amount we can add to any aspect percentage so that the other aspects will still be positive after the redistribution. The median of (-0.07) and 0.47 is 0.2, which we add to get the median of any aspect. Doing this on the seven hazard aspects, 21 sets of weights were produced (Table 3). Each set was used in the GA to determine the optimal arrangement.

This is a 'numerical' effort at understanding the relative importance of the components of vulnerability.

## RESULTS AND DISCUSSIONS

The sensitivity analysis gave two 'best' sets:

- maximizing "Percentage of Population" decreased vulnerability and cost by 23.639 and 13.821 percent respectively.
- maximizing "Size of Affected Area" decreased vulnerability and cost by 1.981 and 35.477 percent, respectively.

Setting the percentage of "Percentage of Population" to maximum produces the highest decrease in vulnerability, while the second best set focused on decreasing the cost. We chose the first set as input to the GA, since it decreased vulnerability by almost 24 %.

The GA produced the following arrangements for the 16 *barangays*, (Figures 3-6*)*:

1. Urbanization: Not surprisingly, the over-all recommendation from both sets of experiments is to make or keep areas in the north (near the river) 'green' or reserving those areas as protected zones. Currently, those are private farm-lands which extend all the way to the river and beyond.

2. Literacy Rate: Currently, the entire city of Urdaneta has a rather high literacy rate, but the recommendation is for less literacy in some areas, possibly as a way of saving money. We interpret it as: we do not need to spend any more on literacy for Urdaneta. Money set for expansion and improvement of education and literacy might be better saved for other needs, like improving health.

3. Mortality Rate: This is where resources could be placed. The recommendation is to improve health services, particularly in areas north, or closer to the river. If we take San Jose (on the west, identified as a flood-prone area) as an example, the recommendation was to decrease literacy rate, but improve on the mortality rate. Here we see some 'realignment' of funds being suggested, e.g., if we have to divert funds, we may take out of literacy, and place in the health programs.

4. Poverty Levels: Figures 3 and 4 show the current state and recommendations for poverty, respectively. There are differing economic levels in the town, and the experiments recommend making some areas less well-off (like the flood-prone Camantiles), while improving other areas (like Pinmaludpod, with relatively neutral physical vulnerability). We interpret this as preferentially locating livelihood projects in Pinmaludpod and other areas whose poverty levels need to be improved.

Additionally, the city government might consider a 'tax' on the *barangays* which are relatively well-off, to help in the development of areas needing economic improvement. This tax needs to be local, as national taxes in the Philippines go to a 'regional fund', which might prove ineffective in 'trickling down' to the municipality.

5. Radio/TV Penetration: Currently, Urdaneta is good in this component, and we do not need to allocate more resources to this; similar to the literacy situation.

6. Nonstructural Measures: The recommendation is for a town-wide improvement in legislation and enforcement of laws concerning flooding. Again not a surprising recommendation, specially for the most densely populated areas like the city-center or Poblacion. A common problem in such areas would be litter in the streets clogging drains, and easements around water-ways.

7. Structural Measures: The recommendation here is to build dikes or flood gates along the river (Figures 5 and 6 show current and recommended profiles, respectively, for structural measures). This will most likely meet with resistance, as the national government might find it hard to justify protecting agricultural lands, but floods damage seedlings, severely limits harvest, and threaten the population.

We recall that the initial analysis placed the highest weight on "structural measures" (6.76), followed by "percent of population" (6.09) and "urbanization" (6.08). After the sensitivity analysis and optimization of percentages and weights, the highest weight is now placed on "percent of population" (8.6), followed by "structural measures" (7.54), and "radio/TV penetration" (6.98), a clear refinement or correction on the model. Recall that "percent of population" is a static chromosome, and the genetic algorithm does not alter it, but the sensitivity analysis showed it to be the most important chromosome when it comes to maximizing/minimizing expectations.

## CONCLUSIONS

We have shown a way of gaining insights on allocation or alignment of resources among the various *barangays* of Urdaneta, to lower its over-all flooding vulnerability. Admittedly, some recommendations might have been expected, some might be difficult to implement. The analysis and specific recommendations for Urdaneta presented here could guide the national government on infrastructure planning, local government on allocation of resources, and the specific areas of concern that can be improved on, with the aim of lowering flooding vulnerability. The analysis could be adopted by national government for other cities, but it should be noted that analysis has to be local, that is, each area has to be studied, and its own vulnerability function calculated, to come up with appropriate recommendations. Future work concerns the cost function, which could be constructed or inferred like 'marginal utility' or 'marginal propensity to consume' in economics, e.g., improvement in poverty alleviation/ money units spent.

**ILLUSTRATIONS AND FIGURES**

Figure 1. Satellite Photo of Urdaneta. Google Earth, July 25, 2015.

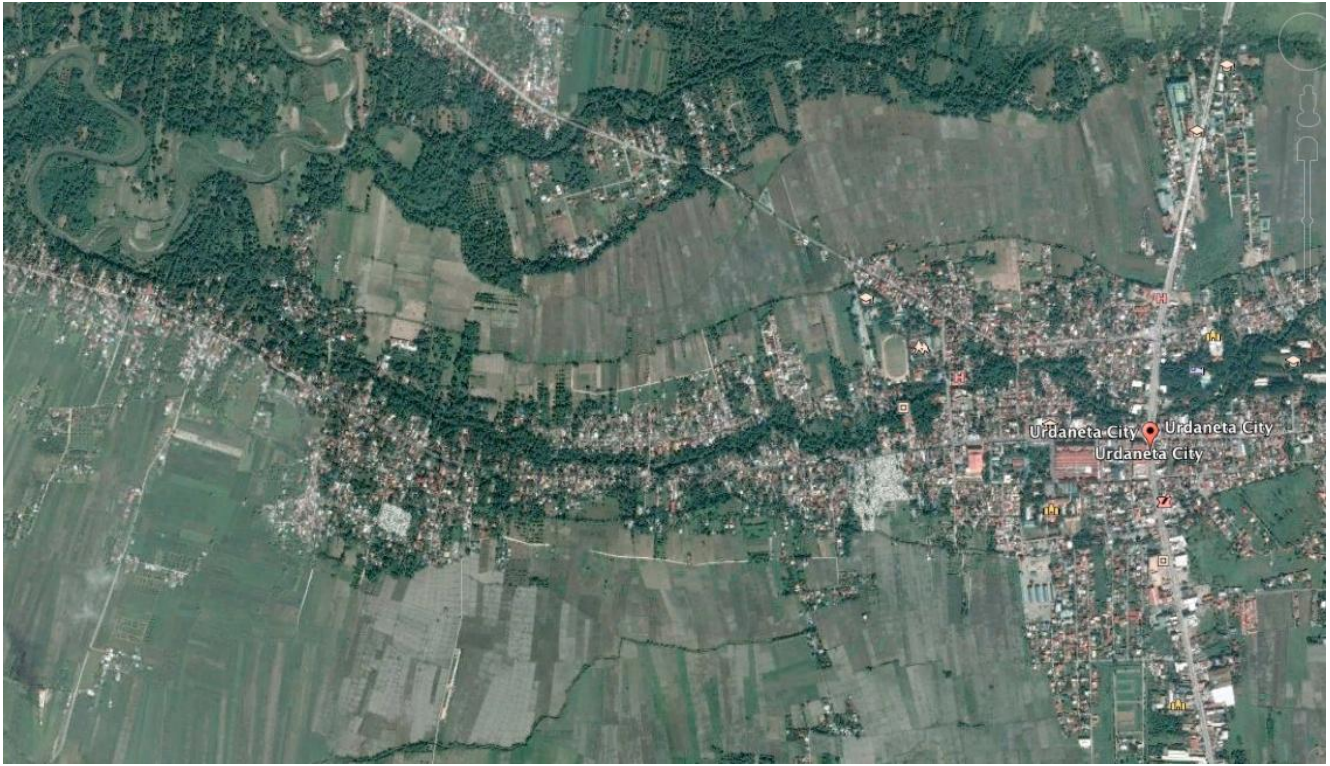

Figure 2. The Physical Area (yellow line indicates the Macalong River)

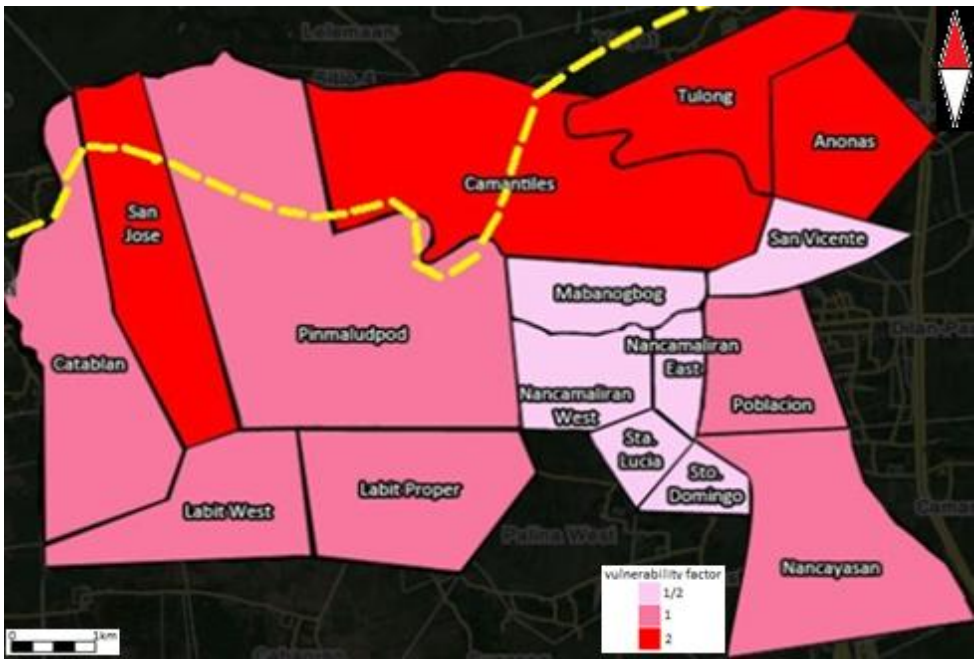

Figure 3. Poverty (Current)

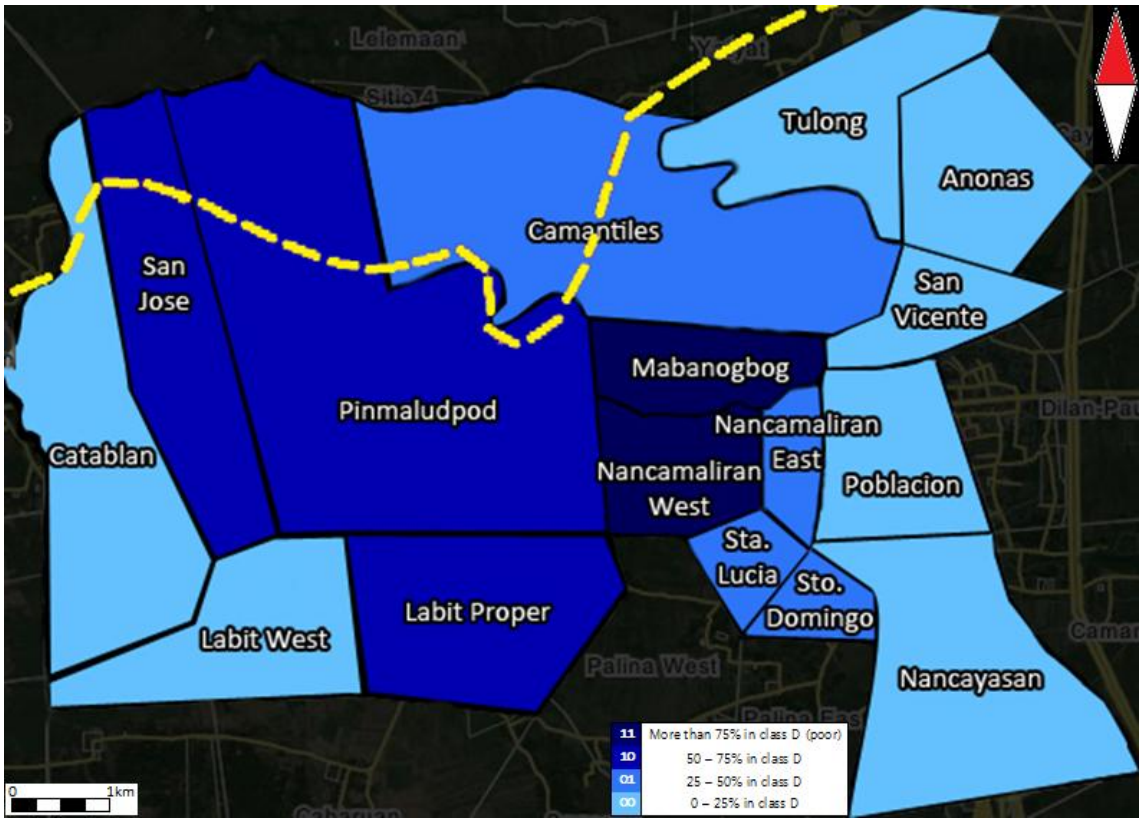

Figure 4. Poverty (Recommended)

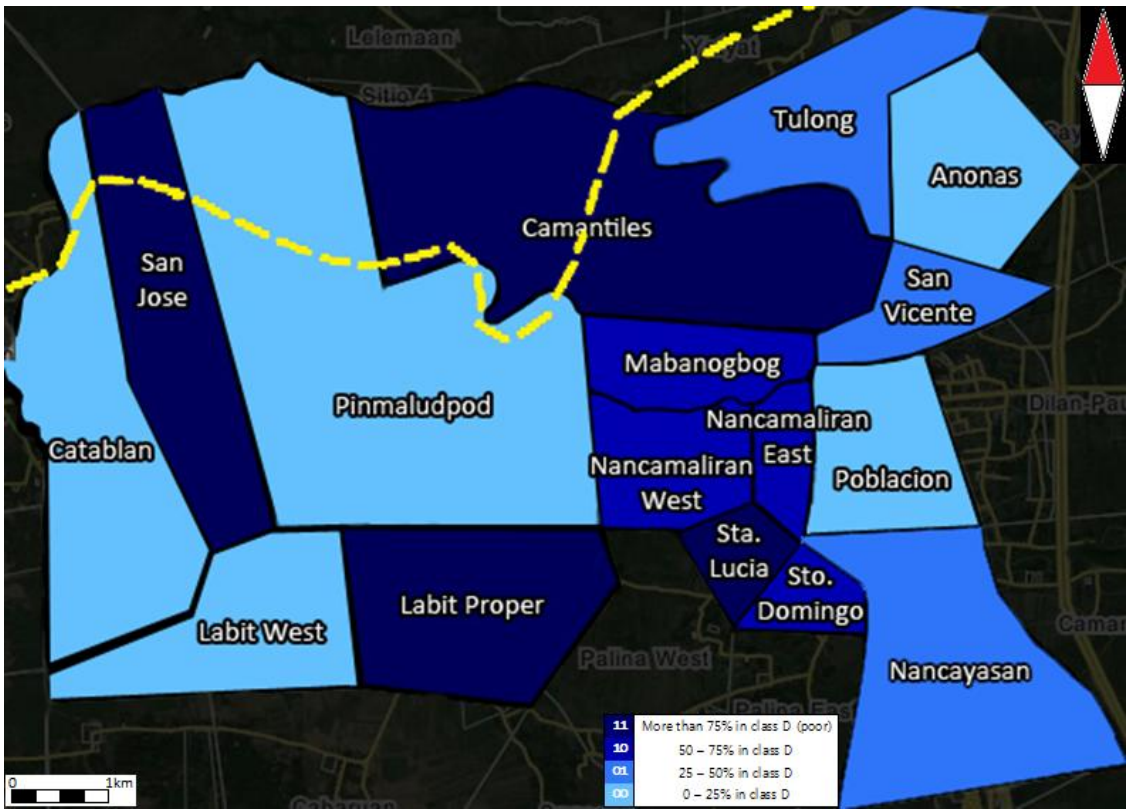

Figure 5. Structural Measures (Current)

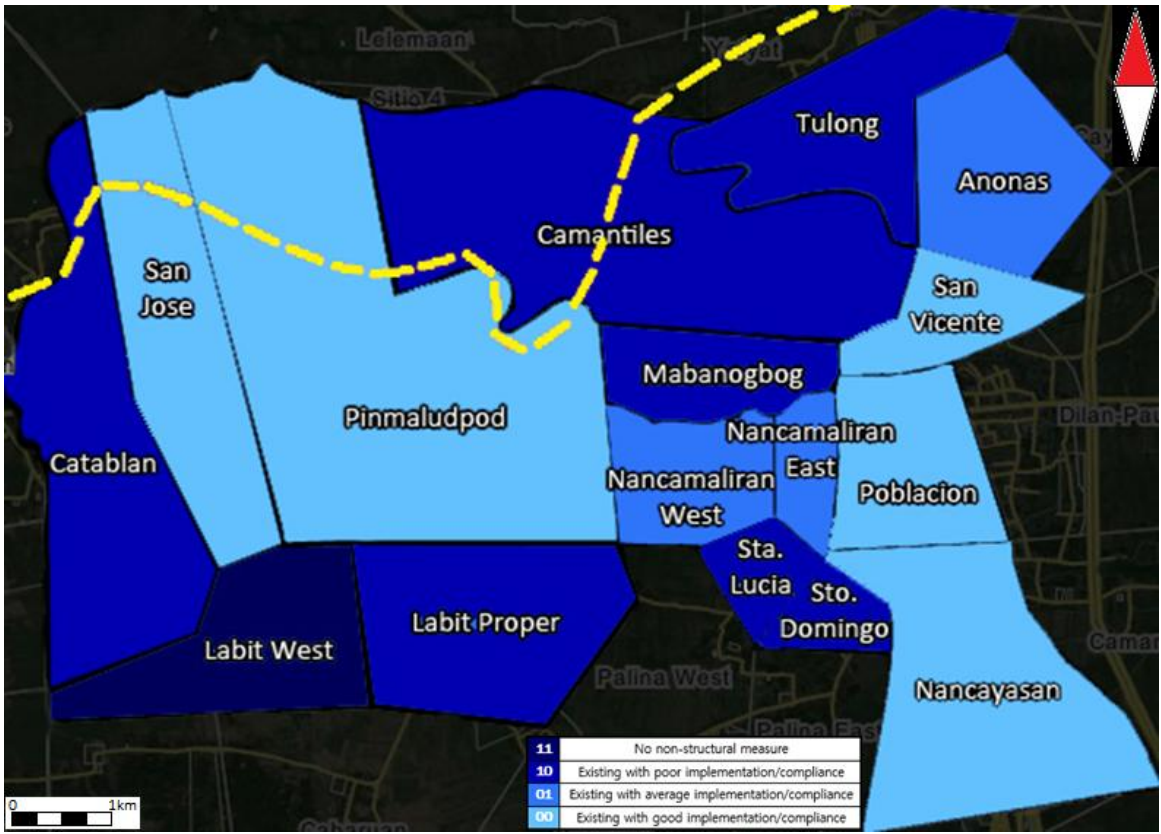

Figure 6. Structural Measures (Recommended).

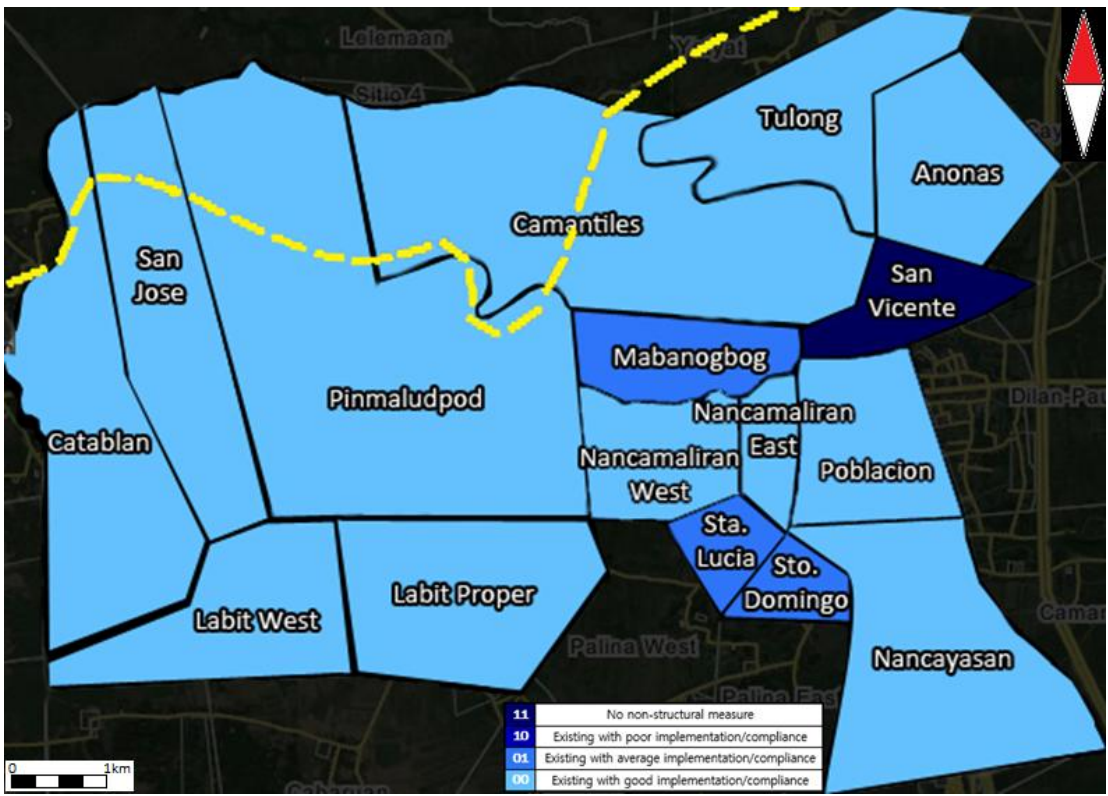

**TABLES AND CAPTIONS**

# Table 1. Vulnerability Determination Table

| Hazard Aspect | Always very important | Usually important | Sometimes important | Rarely of importance | Not worth considering |
|---|---|---|---|---|---|
| Likelihood of occurrence | | | | | ✓ |
| Capacity to cause physical damages | ✓ | | | | |
| Size of affected area | | ✓ | | | |
| Speed of onset (amount of warning time) | | | | ✓ | |
| Percent of population affected | ✓ | | | | |
| Potential for causing casualties | ✓ | | | | |
| Potential for negative economic effects | ✓ | | | | |
| Duration of threat from hazard | | | | ✓ | |
| Seasonal risk pattern | | | | | ✓ |
| Environmental impact | | ✓ | | | |
| Predictability of hazard | | | | ✓ | |
| Ability of hazards to be mitigated | | | | ✓ | |
| Availability of warning systems | | | ✓ | | |
| Public awareness of hazard | | ✓ | | | |
| Corollary effects (ability to cause other hazards) | | | ✓ | | |
| (Other considerations may be added to this list) | | | | | |

## Table 2. Table Analysis

| | Capacity to cause physical damages (19%) | Percentage of the population affected (19%) | Potential for casualties (19%) | Environmental Impact (9%) | Negative economic impacts (18%) | Public awareness of hazard (8%) | Size of Affected Area (8%) | |
|---|---|---|---|---|---|---|---|---|
| Urbanization | 9 -The more urbanized a community is, the more likely for the physical damages to occur. | 7 -More people will be affected if the area is more urbanized. | 6 -If the area is more populated, the potential for casualties would be more likely to increase but not everyone will get injured or die. | 4 -No matter how urbanized a community, the effect of flood is out of our hands. | 5 -More urbanized area will cause less economic loss in the community. | 6 -With an urbanized area, the dissemination of information would be more easier. | 2 -The size of the area is very slightly affected by its urbanization level | (9 x 0.19) + (7 x 0.19) + (6 x 0.19) + (4 x 0.09) + (5 x 0.18) + (6 x 0.08) + (2 x 0.08) = **6.08** |
| Literacy Rate | 3 -If people can understand/read warning signs, they might be more prepared in responding emergency situations as such. | 5 -Lesser percentage of population will be affected since the people are more ready. Although, a number can still be affected for the reason that the effects of the flood are out of their control. | 4 -In an area, a number of people may be affected, but not all are likely to be injured or die, in the worst-case scenario | 2 -People may be able to read and understand warning signs, but the destruction of the environment is out of our hands. | 3 -Knowing the possible effects that the disaster may bring beforehand, certain preventive measures for economic loss can be planned and managed effectively. | 9 -With the capability of the community to read and understand warning signs, they can be well aware of the possible hazards that the flooding can bring. | 2 -Knowing is not enough. People need material resources to be capable of acting effectively. | (3 x 0.19) + (5 x 0.19) + (4 x 0.19) + (2 x 0.09) + (3 x 0.18) + (9 x 0.08) + (2 x 0.08) = **3.88** |
| Mortality Rate | 3 -The number if deaths in a community will not affect much the physical damages that might occur. | 8 -More people will be affected if the mortality rate is high. | 7 -If there is high mortality rate in a community, there is a possibility that casualty will be high. | 2 -The environmental effect is often unpredictable. | 4 -Mortality rate will be a loss that may cause high negative economic impact. | 3 -Being well informed regarding flood and its effect will be of help to the community. | 2 -The size of the affected area does not depend on the mortality rate. | (3 x 0.19) + (8 x 0.19) + (7 x 0.19) + (2 x 0.09) + (4 x 0.18) + (3 x 0.08) + (2 x 0.08) = **4.72** |
| Poverty | 5 -The less fortunate the community, there would be less physical damage that will occur. | 5 -If a community is less fortunate, it will be more likely that the percentage of the population affected will be high. | 4 -Being less fortunate might slightly increase potential for casualty. | 2 -No matter how less fortunate a community is, the effect of flood in the environment is uncontrollable. | 3 -Since the community is less fortunate, there will be less economic impact that will occur | 6 -If they are not equipped with materials that will help them be informed, it will be harder for them to know what is happening. | 2 -Economic status matters little to the size of affected area because of the disaster's magnitude. | (5 x 0.19) + (5 x 0.19) + (4 x 0.19) + (2 x 0.09) + (3 x 0.18) + (6 x 0.08) + (2 x 0.08) = **4.02** |
| TV/Radio Penetration | 6 -The greater the TV and radio penetration, the less likely it would cause physical damages, since certain preventive measures can be implemented. | 8 -There will be a lesser percentage in the population being affected since people are more prepared on how to act in these emergency situations. | 7 -Fewer people are less likely to be injured or die. | 2 -The environmental damage that can be brought by flooding is out of our hands. No matter how aware we can be, the environment can still be greatly affected. | 2 -Through the warnings given via the radio or TV, people may act immediately on their business activities, thereby preventing the decline of economy. | 10 -The public awareness of people prior to the flooding is contributed by the warnings heavily emphasized in TV, radio and other means of communication. | 2 -Awareness may affect the size of affected area. It can only help a small extent. | (6 x 0.19) + (8 x 0.19) + (7 x 0.19) + (2 x 0.09) + (2 x 0.18) + (10 x 0.08) + (2 x 0.08) = **5.49** |

*Table 3 Sets of Weights*

| Hazard Aspect | Operation | M₁ | M₂ | M₃ | M₄ | M₅ | M₆ | M₇ | M₈ | M₉ | M₁₀ | M₁₁ |
|---|---|---|---|---|---|---|---|---|---|---|---|---|
| Original Weights | | 6.08 | 3.88 | 4.72 | 4.02 | 5.49 | 3.78 | 6.76 | 6.09 | 3.29 | 5.43 | 4.86 |
| Capacity to Cause Physical Damages | minimize | 5.8 | 3.96 | 4.81 | 3.93 | 5.43 | 3.86 | 6.56 | 5.88 | 3.3 | 5.27 | 4.74 |
| | median | 6.88 | 3.65 | 4.45 | 4.29 | 5.66 | 3.55 | 7.33 | 6.69 | 3.26 | 5.9 | 5.19 |
| | maximize | 7.96 | 3.33 | 4.09 | 4.65 | 5.88 | 3.23 | 8.09 | 7.5 | 3.21 | 6.53 | 5.64 |
| Percentage of Population | minimize | 5.96 | 3.8 | 4.41 | 3.93 | 5.27 | 3.7 | 6.64 | 5.72 | 3.22 | 5.35 | 4.74 |
| | median | 6.41 | 4.11 | 5.62 | 4.29 | 6.12 | 4.01 | 7.09 | 7.16 | 3.49 | 5.66 | 5.19 |
| | maximize | 6.86 | 4.43 | 6.84 | 4.65 | 6.98 | 4.33 | 7.54 | 8.6 | 3.76 | 5.98 | 5.64 |
| Potential for Casualties | minimize | 6.05 | 3.88 | 4.49 | 4.01 | 5.35 | 3.86 | 6.89 | 6.21 | 3.3 | 5.51 | 4.91 |
| | median | 6.18 | 3.88 | 5.39 | 4.05 | 5.89 | 3.55 | 6.39 | 5.76 | 3.26 | 5.2 | 4.73 |
| | maximize | 6.32 | 3.88 | 6.29 | 4.1 | 6.43 | 3.23 | 5.9 | 5.31 | 3.21 | 4.88 | 4.55 |
| Environmental Impact | minimize | 6.21 | 4.04 | 4.9 | 4.17 | 5.76 | 3.86 | 6.81 | 6.29 | 3.3 | 5.35 | 4.91 |
| | median | 5.71 | 3.41 | 4.22 | 3.59 | 4.72 | 3.55 | 6.63 | 5.52 | 3.26 | 5.66 | 4.73 |
| | maximize | 5.22 | 2.78 | 3.55 | 3 | 3.69 | 3.23 | 6.45 | 4.76 | 3.21 | 5.98 | 4.55 |
| Negative Economic Impacts | minimize | 6.13 | 3.96 | 4.73 | 4.09 | 5.76 | 3.86 | 6.81 | 6.04 | 3.22 | 5.27 | 4.83 |
| | median | 5.95 | 3.65 | 4.69 | 3.82 | 4.72 | 3.55 | 6.63 | 6.22 | 3.49 | 5.9 | 4.96 |
| | maximize | 5.77 | 3.33 | 4.64 | 3.55 | 3.69 | 3.23 | 6.45 | 6.4 | 3.76 | 6.53 | 5.1 |
| Public Awareness of Hazard | minimize | 6.05 | 3.47 | 4.81 | 3.85 | 5.11 | 3.62 | 7.05 | 6.21 | 3.38 | 5.59 | 4.99 |
| | median | 6.18 | 5.05 | 4.45 | 4.52 | 6.59 | 4.25 | 5.93 | 5.76 | 3.02 | 4.96 | 4.49 |
| | maximize | 6.32 | 6.62 | 4.09 | 5.2 | 8.08 | 4.88 | 4.8 | 5.31 | 2.66 | 4.33 | 4 |
| Size of Affected Area | minimize | 6.37 | 4.04 | 4.9 | 4.17 | 5.76 | 3.7 | 6.56 | 6.29 | 3.3 | 5.68 | 4.91 |
| | median | 5.25 | 3.41 | 4.22 | 3.59 | 4.72 | 4.01 | 7.33 | 5.52 | 3.26 | 4.73 | 4.73 |
| | maximize | 4.12 | 2.78 | 3.55 | 3 | 3.69 | 4.33 | 8.09 | 4.76 | 3.21 | 3.79 | 4.55 |

Where:

$M_1$ = Urbanization

$M_2$ = Literacy Rate

$M_3$ = Mortality Rate

$M_4$ = Poverty

$M_5$ = TV/Radio

$M_6$ = Non-Structural

$M_7$ = Structural Measures

$M_8$ = Percentage of Population

$M_9$ = Percentage of Area/Extent

$M_{10}$ = Economic Value

$M_{11}$ = Cost of Relocation